\documentclass[12pt]{article}
\usepackage{html}
\usepackage{epsf}
\usepackage{graphicx}
\usepackage{amssymb}
\begin{document}
\title{MATTERS OF GRAVITY, The newsletter of the APS Topical Group on 
Gravitation}
\begin{center}
{ \Large {\bf MATTERS OF GRAVITY}}\\ 
\bigskip
\hrule
\medskip
{The newsletter of the Topical Group on Gravitation of the American Physical 
Society}\\
\medskip
{\bf Number 39 \hfill Winter 2012}
\end{center}
\begin{flushleft}
\tableofcontents
\vfill\eject
\section*{\noindent  Editor\hfill}
David Garfinkle\\
\smallskip
Department of Physics
Oakland University
Rochester, MI 48309\\
Phone: (248) 370-3411\\
Internet: 
\htmladdnormallink{\protect {\tt{garfinkl-at-oakland.edu}}}
{mailto:garfinkl@oakland.edu}\\
WWW: \htmladdnormallink
{\protect {\tt{http://www.oakland.edu/?id=10223\&sid=249\#garfinkle}}}
{http://www.oakland.edu/?id=10223&sid=249\#garfinkle}\\

\section*{\noindent  Associate Editor\hfill}
Greg Comer\\
\smallskip
Department of Physics and Center for Fluids at All Scales,\\
St. Louis University,
St. Louis, MO 63103\\
Phone: (314) 977-8432\\
Internet:
\htmladdnormallink{\protect {\tt{comergl-at-slu.edu}}}
{mailto:comergl@slu.edu}\\
WWW: \htmladdnormallink{\protect {\tt{http://www.slu.edu/colleges/AS/physics/profs/comer.html}}}
{http://www.slu.edu//colleges/AS/physics/profs/comer.html}\\
\bigskip
\hfill ISSN: 1527-3431

\bigskip

DISCLAIMER: The opinions expressed in the articles of this newsletter represent
the views of the authors and are not necessarily the views of APS.
The articles in this newsletter are not peer reviewed.

\begin{rawhtml}
<P>
<BR><HR><P>
\end{rawhtml}
\end{flushleft}
\pagebreak
\section*{Editorial}

The next newsletter is due September 1st.  This and all subsequent
issues will be available on the web at
\htmladdnormallink 
{\protect {\tt {https://files.oakland.edu/users/garfinkl/web/mog/}}}
{https://files.oakland.edu/users/garfinkl/web/mog/} 
All issues before number {\bf 28} are available at
\htmladdnormallink {\protect {\tt {http://www.phys.lsu.edu/mog}}}
{http://www.phys.lsu.edu/mog}

Any ideas for topics
that should be covered by the newsletter, should be emailed to me, or 
Greg Comer, or
the relevant correspondent.  Any comments/questions/complaints
about the newsletter should be emailed to me.

A hardcopy of the newsletter is distributed free of charge to the
members of the APS Topical Group on Gravitation upon request (the
default distribution form is via the web) to the secretary of the
Topical Group.  It is considered a lack of etiquette to ask me to mail
you hard copies of the newsletter unless you have exhausted all your
resources to get your copy otherwise.

\hfill David Garfinkle 

\bigbreak

\vspace{-0.8cm}
\parskip=0pt
\section*{Correspondents of Matters of Gravity}
\begin{itemize}
\setlength{\itemsep}{-5pt}
\setlength{\parsep}{0pt}
\item John Friedman and Kip Thorne: Relativistic Astrophysics,
\item Bei-Lok Hu: Quantum Cosmology and Related Topics
\item Veronika Hubeny: String Theory
\item Pedro Marronetti: News from NSF
\item Luis Lehner: Numerical Relativity
\item Jim Isenberg: Mathematical Relativity
\item Katherine Freese: Cosmology
\item Lee Smolin: Quantum Gravity
\item Cliff Will: Confrontation of Theory with Experiment
\item Peter Bender: Space Experiments
\item Jens Gundlach: Laboratory Experiments
\item Warren Johnson: Resonant Mass Gravitational Wave Detectors
\item David Shoemaker: LIGO Project
\item Stan Whitcomb: Gravitational Wave detection
\item Peter Saulson and Jorge Pullin: former editors, correspondents at large.
\end{itemize}
\section*{Topical Group in Gravitation (GGR) Authorities}
Chair: Patrick Brady; Chair-Elect: 
Manuela Campanelli; Vice-Chair: Daniel Holz. 
Secretary-Treasurer: James Isenberg; Past Chair:  Steve Detweiler;
Members-at-large:
Scott Hughes, Bernard Whiting,
Laura Cadonati, Luis Lehner,
Michael Landry, Nicolas Yunes.
\parskip=10pt

\vfill
\eject

\vfill\eject

\section*{\centerline
{News from NSF}}
\addtocontents{toc}{\protect\medskip}
\addtocontents{toc}{\bf GGR News:}
\addcontentsline{toc}{subsubsection}{
\it News from NSF, by David Garfinkle}
\parskip=3pt
\begin{center}
David Garfinkle
\htmladdnormallink{garfinkl-at-oakland.edu}
{mailto:garfinkl@oakland.edu}
\end{center}

Beverly Berger retired as Program Director for Gravitational Physics at the end of 2012. The new program director is Dr. Pedro Marronetti from Florida Atlantic University. Beverly wishes to thank the gravitational physics community for their exciting discoveries and results over the past decade and wishes everyone even greater future successes. She also wishes to thank everyone who served as a reviewer or panelist for this essential contribution to the health of the NSF Gravity Program.

\vskip1.0truein

\section*{\centerline
{we hear that \dots}}
\addtocontents{toc}{\protect\medskip}
\addcontentsline{toc}{subsubsection}{
\it we hear that \dots , by David Garfinkle}
\parskip=3pt
\begin{center}
David Garfinkle, Oakland University
\htmladdnormallink{garfinkl-at-oakland.edu}
{mailto:garfinkl@oakland.edu}
\end{center}

Jacob Bekenstein has received the Wolf Prize.

Carl Brans, Alessandra Buonanno, Curt Cutler, Frans Pretorius, Oscar Reula, and Alexei Starobinsky 
have been elected APS Fellows.  

Hearty Congratulations!

\vfill\eject

\section*{\centerline
{GGR program at the APS meeting in Atlanta, GA}}
\addtocontents{toc}{\protect\medskip}
\addcontentsline{toc}{subsubsection}{
\it GGR program at the APS meeting in Atlanta, GA, by David Garfinkle}
\parskip=3pt
\begin{center}
David Garfinkle, Oakland University
\htmladdnormallink{garfinkl-at-oakland.edu}
{mailto:garfinkl@oakland.edu}
\end{center}

We have an exciting GGR related program at the upcoming APS April meeting in Atlanta, GA.  
Our Chair-Elect, Manuela Campanelli did an 
excellent job of putting together this program.  At the APS meeting there will be several invited sessions of talks sponsored
by the Topical Group in Gravitation (GGR).

The invited sessions sponsored by GGR are as follows:\\

Numerical Relativity Beyond Astrophysics\\
Luis Lehner, Higher dimensional gravity and black holes\\
Paul Chesler, Gravity in asymptotically Anti de Sitter (AdS) and AdS/CFT\\
Ulrich Sperhake, High speed black hole collisions with applications to trans-Planckian particle scattering\\

Analytical Relativity Meets Numerical Relativity\\
Ajith Parameswaran, Interfacing analytical and numerical relativity for gravitational wave astronomy\\
Alexandre Le Tiec, The overlap of numerical relativity, perturbation theory and post-Newtonian theory in the binary black hole problem\\
Ryan Lang, Binary black hole mergers: astrophysics and implications for space based gravitational wave detectors\\ 

Pulsar Timing Arrays and Gravitational Radiation\\
(Joint with DAP)\\
Andrea Lommen, Pulsar timing arrays: no longer a blunt instrument for gravitational wave detection\\
Zoltan Haiman, Electromagnetic emission from supermassive black hole binaries resolved by pulsar timing arrays\\
Massimo Tinto, Testing alternative theories of gravity using pulsar timing arrays\\

Challenges in Numerical Relativity\\
(joint with COM)\\
Pablo Laguna, Numerical relativity and black hole binaries: the historical path to present simulations\\
Carlos Palenzuela, Computing EM signatures from astrophysical compact binary mergers\\
Pedro Marronetti, Simulations of core collapse supernovae\\

Advances in Quantum Gravity\\
Diego Hofman, Applications of holography to condensed matter physics\\
Ivan Agullo, Beyond the standard inflationary paradigm\\
Eugenio Bianchi, Loop quantum gravity, spin foams, and gravitons\\

Testing Gravity on Cosmic Scales\\
(joint with DAP)\\
Bhuvnesh Jain, Astrophysical tests of gravity\\
Claudia deRham, Theoretical developments in bounding gravity in extra dimensions\\
Fabian Schmidt, N-body simulations of modified gravity models\\

Perspectives in Gravitational Physics\\
Beverly Berger, Perspectives in Gravitational Physics\\
Kip Thorne, Perspectives on geometrodynamics: the nonlinear dynamics of curved spacetime\\
John Friedman, Perspectives on relativistic astrophysics in the century's first decade\\

The Detection Challenge in Modern Physics Experiments\\
(joint with DPF)\\
Gabriela Gonzalez, Gravitational wave astronomy with LIGO and Virgo detectors\\
Dario Autiero, Results from OPERA on superluminal neutrinos\\
Gray Rybka, ADMX: the Axion Dark Matter eXperiment\\

The GGR contributed sessions are as follows:\\

Current and Future Gravitational Wave Experiments\\

Gravitational-Wave Source Modeling\\

Gravitational Waves: Data Analysis\\
(joint with DAP)

General Relativistic Magnetohydrodynamical Simulations\\
(joint with DCOMP)

Numerical Relativity: Neutron Stars and Black Holes\\

Simulations of General Relativistic Astrophysical Phenomena\\

Tests of General Relativity and Gravitation Sponsor\\

Quantum Aspects of Gravitation\\

Numerical Relativity: Black-Hole Binaries\\

Extracting Physical Information from Gravitational Wave Signals\\
(joint with DCOMP)

Approximations in General Relativity\\

Exact Solutions and Analyses of Spacetimes\\

Topics in Numerical Relativity\\

General Relativity and Cosmology\\

\section*{\centerline
{AdS instability}}
\addtocontents{toc}{\protect\medskip}
\addtocontents{toc}{\bf Research briefs:}
\addcontentsline{toc}{subsubsection}{
\it AdS instability, by David Garfinkle}
\parskip=3pt
\begin{center}
David Garfinkle, Oakland University
\htmladdnormallink{garfinkl-at-oakland.edu}
{mailto:garfinkl@oakland.edu}
\end{center}

\begin{figure}
\includegraphics{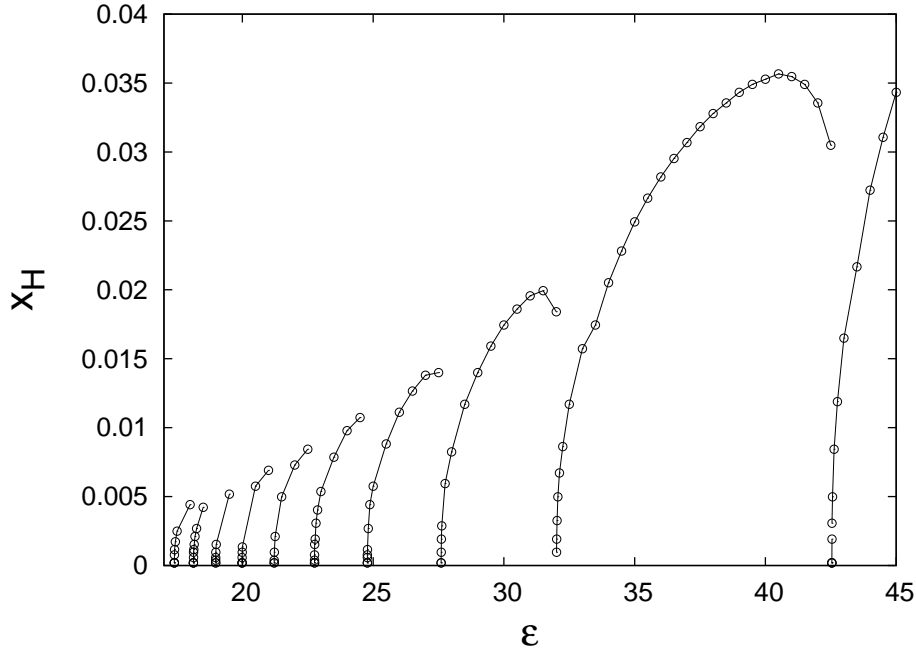}
\caption{critical collapse in AdS
\label{AdS1}}
\end{figure}

It is well known that Minkowski spacetime is nonlinearly stable\cite{christodoulou} in the sense that 
sufficiently small perturbations
remain small.  The same can also be shown for de Sitter spacetime.\cite{friedrich}  But no such result has been 
shown for anti de Sitter (AdS) spacetime.  Recently a remarkable combination of analytical and numerical 
results\cite{bizon1} shows that AdS is nonlinearly unstable.  The results of \cite{bizon1} begin
with a numerical study of critical gravitational collapse in spherically symmetric spacetimes that are asymptotically 
AdS and have a scalar field as the matter.  Recall that in
numerically studying critical collapse one prepares a one parameter family of initial data and looks for scaling of 
black hole mass at the threshold of black hole formation.  The results of this study are shown in fig. (\ref{AdS1}) from
\cite{bizon1} which is a plot of black hole horizon size as a function of the amplitude of the initial scalar field.  
The rightmost curve in the plot is the sort of behavior one obtains in the asymptotically flat case: as one approaches the 
critical parameter, black hole mass approaches zero as a power law.  However, what happens to the left of that curve is completely different: in the asymptotically flat case the ingoing waves that are just barely too weak to form a black hole  disperse to infinity.  But in the AdS case, such wave packets reach AdS infinity 
in a finite time, bounce back and then form a black hole 
on the second try.  Those that just barely don't form a black hole on the second try bounce off AdS infinity again and form a black hole on the third try, and so on, thus leading to all the curves in the figure.

\begin{figure}
\includegraphics{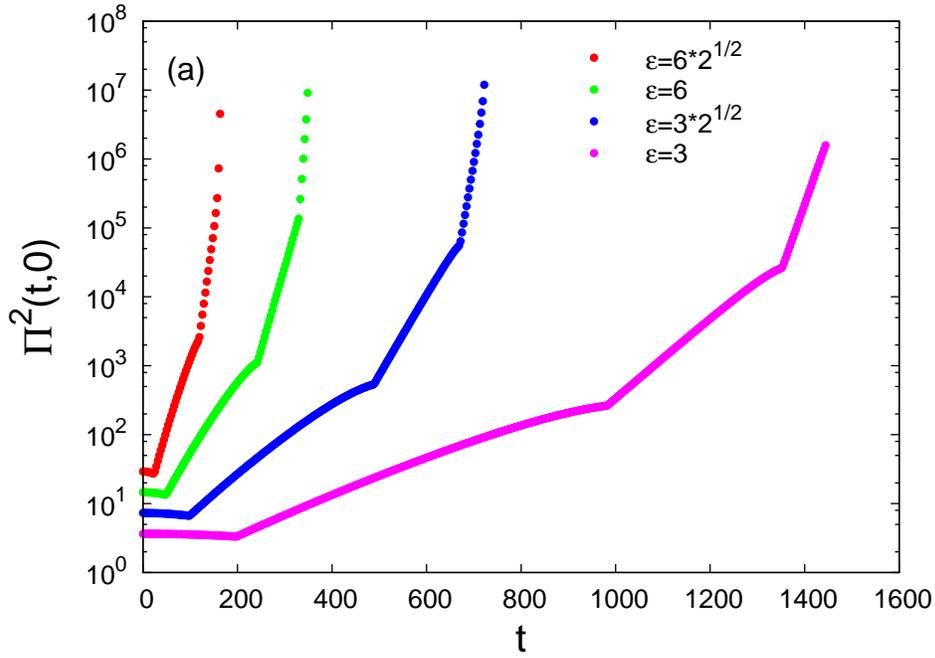}
\caption{onset of the nonlinear instability
\label{AdS2}}
\end{figure}

What accounts for this difference in behavior between the asymptotically flat case and the asymptotically AdS case?  Certainly a large part of the difference is that AdS infinity acts like the walls of a box and keeps the scalar field from dispersing.  But failure to disperse does not imply black hole formation.  Rather black hole formation requires a steepening of the scalar field wave packet.  That is, the fixed energy of the scalar field must be compressed into a smaller volume.  To account for the steepening, the authors of \cite{bizon1} perform a calculation in perturbation theory.  They note that in certain cases the nonlinear effects of two zeroth order modes is to give rise to another mode of higher frequency.  These higher frequency modes in turn can combine with other modes to make modes of even higher frequency, and so on.  The net effect is to put more energy into higher frequencies and thus lead to an ever steepening wave packet.  This effect is illustrated in fig. (\ref{AdS2}) from \cite{bizon1} which plots the time derivative of the scalar field at $r=0$ vs time.  These results are from simulations of the evolution of initial data that begin with small amplitude, and for clarity only the upper envelope of the rapidly oscillating scalar field is plotted.       

From the point of view of general relativity, spacetimes of dimension 4 (which is the case treated in \cite{bizon1}) 
is the most interesting case.  However, asymptotically AdS spacetimes are also of interest due to the AdS/CFT 
correspondence\cite{maldacena}: a conjectured duality between asymptotically AdS spacetimes of dimension $n+1$ and conformal field theories on the dimension $n$ AdS infinity boundary of those spacetimes.  From the AdS/CFT point of view the most 
interesting spacetime dimension is 5 since it yields information about 4 dimensional quantum field theory.  One is thus
led to the quesion of whether the results of \cite{bizon1} generalize to the case of dimension 5.  This question was
answered in the affirmative in \cite{bizon2}.  Here it was shown that both the numerical results and the analytical perturbative
results continue to hold in dimension 5.  That is, arbitrarily small initial data undergo multiple bounces, steepen, and eventually form black holes.  At first this might seem puzzling from the point of view of AdS/CFT since black holes in the spacetime correspond to thermal states in the field theory.  Thus the results of \cite{bizon2} imply that in general states of the field theory thermalize.  However, it is important to remember that in the boundary spacetime, space is a three-sphere. 
Thus, it is not surprising that quantum fields confined to the finite volume of a three-sphere will eventually thermaize.

Since the results of \cite{bizon1,bizon2} assume spherical symmetry, it is natural to ask whether they generalize to the 
case without symmetry.  This question was addressed in\cite{gary1}.  Here a perturbative treatment was performed of vacuum
(with a cosmological constant) spacetimes without symmetry, and that are asymptotically AdS.  Here it was shown that the
same phenomena are found that occur in the perturbative treatment of \cite{bizon1}.  That is small initial data with multiple
modes give rise to modes of higher frequency through the nonlinearities of the Einstein field equation.  Though beyond the scope
of the perturbative treatment of \cite{gary1} to prove, it is nonetheless convincingly argued that this steepening will eventually lead to the formation of a black hole and therefore that vacuum AdS is unstable to black hole formation. 

What then is the endstate of a slightly perturbed AdS spacetime?  One might expect that the perturbation evolves to an
asymptotically AdS black hole as the endstate.  However, it has been argued\cite{gary2} that asymptotically AdS black holes
are also unstable.  It is thus an open question as to what the endstate might be.

\vfill\eject

\section*{\centerline
{Isenberg Fest}}
\addtocontents{toc}{\protect\medskip}
\addtocontents{toc}{\bf Conference reports:}
\addcontentsline{toc}{subsubsection}{
\it Isenberg Fest, 
by Robert M. Wald}
\parskip=3pt
\begin{center}
Robert M. Wald, University of Chicago 
\htmladdnormallink{rmwa@uchicago.edu}
{mailto:rmwa@uchicago.edu}
\end{center}

Jim Isenberg's 60th birthday was celebrated at the Fall Meeting of the Pacific Northwest Geometry Seminar, held at Oregon State University on the weekend of November 12-13, 2011.

The meeting began with a reception and banquet in Jim's honor on the evening of Friday, November 11. Jim is a rather colorful figure, so there was no shortage of toasts involving topics such as marathons/running, Spanish peanuts, and the Boston Red Sox. But the most frequently mentioned topic was Jim's farm, where he raises such animals as goats, sheep, and llamas. Those of Jim's collaborators who accept his generous
hospitality and stay at his farm are often invited to help out with some farm tasks, such as bringing a buck down a muddy
hillside in the rain. (Hillsides in Oregon are always muddy, because, as I learned, in Oregon ``drought conditions" means "drizzle".) One of the conference organizers, Dan Knopf, succeeded in declining such an invitation from Jim. However, his observations of Jim over the course of the subsequent hour have provided conclusive proof that, in muddy conditions,
``four-hoof drive" is far superior to ``two-leg drive" (particularly with bad knees). The evening was literally ``capped off" with the sale of ``Jim Is Old" baseball hats,
complete with a goat insignia.

The talks on Saturday and Sunday provided a very interesting mix of geometry and mathematical general relativity. This mix of topics was highly appropriate in view of Jim's interests, and, indeed, talks were given by Jim's collaborators both on the geometry side (Mazzeo and Knopf) and the general relativity side (Moncrief, Berger, and Chrusciel). The schedule also allowed ample time for informal discussions among the ~100 attendees. The speakers and their titles were:

     Rafe Mazzeo (Stanford), ``Canonical metrics on singular spaces"

     Vincent Moncrief (Yale), ``Are Analytic Compact Cauchy Horizons Necessarily Killing Horizons?"

     Beverly Berger (NSF), ``The Mysterious Case of Expanding Galileo Spacetimes

     Dan Knopf (U of Texas at Austin) ``Neckpinch dynamics for asymmetric surfaces evolving by mean curvature flow"

     Rick Schoen (Stanford) ``A new mean curvature proof of the spacetime positive mass theorem"

     Piotr Chruściel (University of Vienna) ``Lorentzian geometry with continuous metrics"

     Gary Horowitz (UCSB Physics) ``Instability of Anti-de Sitter Spacetime"

Excellent videos of all of these talks (as well as abstracts and, in some cases, a list of open problems associated with the talks) can be found at:
\htmladdnormallink 
{\protect {\tt {http://www.math.washington.edu/$\sim$lee/PNGS/2011-fall/}}}
{http://www.math.washington.edu/~lee/PNGS/2011-fall/} 
     
There was one additional major highlight of the weekend, and Jim would surely consider me to be remiss if I did not report it here: On Saturday night, the Oregon Ducks upset Stanford 53-30.

\vfill\eject

\section*{\centerline
{COSMO 11}}
\addtocontents{toc}{\protect\medskip}
\addcontentsline{toc}{subsubsection}{
\it COSMO 11, 
by Carlos Martins}
\parskip=3pt
\begin{center}
Carlos Martins, University of Porto
\htmladdnormallink{Carlos.Martins-at-astro.up.pt}
{mailto:Carlos.Martins@astro.up.pt}
\end{center}

The fifteenth edition of the annual International Conference on Particle Physics and Cosmology - COSMO 11 - was held in Porto (often known to foreigners as Oporto), Portugal, in the week of August 22-26, 2011. It was hosted by the Center for Astrophysics of the University of Porto (CAUP, the leading Portuguese research unit in the field), and co-organized by four other research units (the University of Porto's CFP, the University of Lisbon's CAAUL and CFNUL, and the Technical University of Lisbon's CENTRA).

Starting as a workshop in Ambleside in 1997, COSMO meetings rapidly became the key forum where particle physicists, cosmologists and astrophysicists worldwide meet and exchange ideas in one of the most active areas in physics. The entire conference was recorded and livestreamed (with the exception of 10 speakers who declined to be filmed), and in addition to the 244 in situ participants we estimate that around 100 other people watched the webcast of some of the talks.

The conference was devoted to the new and exciting interfaces between Fundamental and Phenomenological Particle Physics and Physical Cosmology and Astronomy. There were a total of 20 plenary talks, 102 parallel session talks, and 53 posters. As soon as the conference proceedings DVD is released and sent to participants, an online version of it (including all the recorded talks and the pdfs of all oral contributions and posters) will be available online at 
\htmladdnormallink {\protect {\tt {http://www.astro.up.pt/cosmo11}}}
{http://www.astro.up.pt/cosmo11}

The plenary review talks can be broadly grouped into six different topics

- Standard cosmology: Licia Verde reviewed the status of cosmological parameters determination and showed some examples of testing deviations from the standard cosmological model. Karsten Jedamzik discussed the current status of big bang nucleosynthesis, particularly the Lithium problem and possible astrophysical and beyond the standard model solutions to it, and also presented a new BBN code. Antonio Riotto presented an overview of the phenomenology and observational status of inflation, with particular emphasis on the theoretical challenges we might be facing the next few years. Julien Lesgourgues reviewed the cosmological constraints on neutrino properties and discussed progress in modelling and computing the impact of neutrinos on cosmological observables, including in the non-linear regime.

- Dark matter theory: Katie Freese reviewed theoretical dark matter candidates, focusing on WIMPs, and also discussed the scenario of dark stars. Ki Young Choi talked about the candidates of dark matter beyond standard WIMPs, especially gravitino and axino dark matter and its relation to the early Universe and collider experiments. Yvonne Wong reviewed the role of the QCD axion in cosmology as a potential dark matter candidate, and discussed current cosmological and astrophysical constraints on axions and axion-like particles.

- Dark matter experiments: Antonio Melgarejo discussed the challenges of direct dark matter searches and the latest results of several ongoing direct dark matter experiments. Pierre Salati reviewed the current experimental and theoretical searches for the indirect presence of dark matter species inside the Milky Way halo, discussing in detail the astrophysical backgrounds inside which the various signals are hidden. Simon White discussed the present distribution of dark matter halos, up to the extremely nonlinear regions such as the part of the Galaxy where the Sun resides.

- Extra dimensions: Liam McAllister discussed recent attempts to realize inflation (or an alternative to it) in string theory. Ana Achucarro discussed the effects of very heavy fields present during inflation, which are generic in Supergravity and Superstring models. Cliff Burgess summarized recent progress on understanding back-reaction for higher codimension branes, together with some interesting new implications they can have for particle physics and cosmology. Howard Baer reviewed the latest searches for SUSY from Atlas and CMS, and what these imply for supersymmetric theories and the dark matter of the universe.

- Dark energy: Federico Piazza highlighted the role of the Equivalence Principle for modelling and testing physics beyond General Relativity and the Standard Model, and outlined its present experimental status. Jean-Philippe Uzan reviewed the progress in testing the equivalence principle on astrophysical scales, particularly using fundamental constants, and also discussed the link with dark energy models. Ed Copeland reviewed the zoo of available dynamical dark energy models, and discussed possible ways of distinguishing between them.

- New signatures: Juan Garcia-Bellido discussed the present status of the theory of preheating and reheating after inflation, focusing on the phenomenological signatures we may detect in the future from a gravitational wave background, cosmological magnetic fields and baryogenesis. Ruth Durrer reviewed the possibilities to generate the cosmological seeds of the magnetic fields observed in galaxies and clusters, and the possibilities for detecting them in the CMB or the gravitational wave background. Paul Shellard reviewed prospects for uncovering non-Gaussian signatures of new physics from the early universe, describing in particular new modal methods for efficient and optimal extraction of higher order correlators from large datasets.

We had originally envisaged six different parallel sessions, each of which had two appointed co-chairs (in charge of selecting the oral communications from among the submitted abstracts and of chairing the corresponding sessions), on the following topics:

- String \& brane cosmology (Philippe Brax and Takahiro Tanaka)
- Inflation \& phase transitions (Arttu Rajantie and Mikhail Shaposhnikov)
- Dark matter \& other relics (Genevieve Belanger and Leszek Roszkowski)
- Dark energy \& modified gravity (Ruth Lazkoz and David Mota)
- Probing particle physics with CMB and LSS (Pedro Ferreira and Alessandro Melchiorri)
- Future probes of fundamental physics (Bruce Bassett and Graca Rocha)

However, after the closing of the abstract submission period we decided to create a seventh session on Non-gaussianities (and reshuffled the session chairs, with Takahiro Tanaka being in charge of it). The large number of abstracts received on this topic (most of them having PhD students or young post-docs as first authors) highlights it as a particularly active topic of current research.

A novelty in COSMO meetings was that we had a poster competition. All registered participants could vote on the best poster, and in parallel there was a second (independent) vote by the members of the COSMO Steering Committee in attendance. Active lobbying could be seen during the poster session and coffee breaks, and by combining the results of the two votes three winners were selected and announced in the closing session: Danielle Wills, Florian Kuhnel and Guido Walter Pettinari. It goes without saying that the prizes were nice bottles of Port wine.

CAUP is a unique research unit in that it shares a building with a planetarium, has four dedicated outreach staff, and CAUP researchers are supposed to devote 5 \% of their time planning, helping with or taking part in outreach activities. Accordingly, we also organized a cycle of four COSMO-11 public talks. Two of these took place in the week before COSMO (given by myself and by Chuck Bennett) and two during the COSMO week itself (given by Jean-Philippe Uzan and Paul Shellard). All were quite well attended, considering that they took place at the peak of the holiday season in Portugal and that three of them were given in English.

Although rain is part of the COSMO traditions and duly appeared (unusually for Porto in August), it was thankfully absent from the free afternoon, which consisted of a short tour of the Porto town (including a cruise along the Douro river) and a dinner at one of the Port wine cellars. The rest of the social programme consisted of an informal reception on the arrival day, and another one to coincide with the poster session. There were also two special planetarium sessions for COSMO participants and their guests.

COSMO-11 also had a satellite meeting: an observational cosmology summer school held at the historic town of Angra do Heroismo in Terceira (one of the Azores Islands) from August 31 to September 6. This was probably the first major international scientific meeting in this field with a majority of women: 23 of the 40 students (and 29 of the 55 total participants) were women, coming from universities in five continents. Video recordings of the lectures will also be made available soon on the school website 
\htmladdnormallink {\protect {\tt {http://www.astro.up.pt/azores11}}}
{http://www.astro.up.pt/azores11}

COSMO-11 had the financial support of the ON.2, a Novo Norte/QREN programme (funded through FEDER, the European Regional Development Fund), the Portuguese research council (FCT), the University of Porto and the Orient Foundation, as well as the organizing research units, and its success was in large part due to the dedication and hard work of the CAUP staff (Elsa Silva, Manuel Monteiro and Paulo Peixoto) and the CAUP undergraduate students (Amelia Mafalda Leite, Ana Catarina Leite, Gil Marques, Mafalda Monteiro, Miguel Oliveira, Pedro Pedrosa and Jose Pedro Vieira).

COSMO-12 will be held in Beijing, China, in the week of September 10-14, and information on it can already be found at 
\htmladdnormallink {\protect {\tt {http://lss.bao.ac.cn/cosmo12/}}}
{http://lss.bao.ac.cn/cosmo12/}
	
\vfill\eject

\section*{\centerline
{Astro-GR 6 2011}}
\addtocontents{toc}{\protect\medskip}
\addcontentsline{toc}{subsubsection}{
\it Astro-GR 6 2011, 
by  Sascha Husa et al}
\parskip=3pt
\begin{center}
Sascha Husa, Sara Gil, Alicia Sintes (University of the Balearic Islands) and Pau Amaro Seoane (Albert Einstein Institute)
\htmladdnormallink{sascha.husa@uni.es}
{mailto:sascha.husa@uni.es}
\end{center}					

Astro-GR is a yearly informal international meeting  on topics of gravitational wave astronomy, in particular with space-based detectors. The last and sixth installment of the series was focused on the impact and scientific potential of the future space interferometer eLISA/NGO, and was hosted by the University of the Balearic Islands in Palma de Mallorca during 5 - 9 September, 2011
\cite{astrogr-web}. More than 80 scientists attended, including Oliver Jennrich, the ESA project scientist. Astro-GR 2011 was run in the style of all previous workshops, with presentations in the morning and discussion/working groups in the afternoon. The working groups were Binary Black Holes (BBH), Extreme-Mass-Ratio Inspiral (EMRI) and Galactic Binaries, and Intermediate-Mass-Ratio Inspirals (IMRI) and Electromagnetic Counterparts. 

Astro-GR 2011 was the first ``LISA''-related meeting after NASA and ESA announced the end to their ten-year-long partnership in this mission. As a consequence, ESA is now developing a redesigned European-only gravitational wave detector in space , currently under the names eLISA/NGO, that could be launched before 2022 \cite{eLISA-web,eLISA-paper}. A new European science team already completed a science performance study, and an industrial study is currently in progress. After studying several configurations, a new baseline has been identified that simplifies the design of LISA, reducing the distance between the satellites and employing only four instead of previously six laser links. ESA will decide this year whether NGO will go forward as a Cosmic Vision L-class mission, in which case NASA may still participate as a partner with a smaller involvement.

All the talks were recorded, and videos and slides are available on the workshop website \cite{astrogr-web}.  Antoine Petiteau, one of the coordinators of the eLISA ``Science Performance Task Force'',  opened the meeting with a presentation of the ``Performance of the new space based gravitational wave detector'', discussing the new baseline consisting of one mother and 2 daughter spacecraft with only 4 (instead of 6 for LISA) laser links and a reduced arm length of one million km.
Antoine discussed the noise budget of the new baseline and alternative configurations, and expected numbers and parameter estimation errors for different types of sources. One significant change in the noise budget would be that galactic binary confusion noise would be almost negligible for the new baseline.
Antoine's talk was complemented by Oliver Jennrich and Michele Vallisneri, reporting the current status of space based GW missions from the ESA and NASA perspectives. To update us on other future dectors, Michele Punturo and B. Sathyaprakash reported on the Einstein Telescope Project; and Neil Cornish asked ``What can we learn about Blacks Holes using Pulsar Timing Arrays?''

A key question regarding the new LISA design - how it will change the prospects for observing EMRIs - was discussed by Jonathan Gair in his talk "Prospects for EMRI detection and science using LISA-light''. Considering different mission configurations, and looking in detail at issues such as spin effects, testing the no hair property, or measuring the Hubble constant ($\approx 1-2\%$ should be possible with about 20 events),  Jonathan finds that prospects for detection of EMRIs with eLISA are still good, and we can expect  $\approx 100$ events at redshift $z < 0.5$ for the new baseline mission.  Descoped missions will offer the same range of science as Classic LISA, albeit with fewer events and reduced SNR for individual sources. However, large rate uncertainties arise from astrophysics.

A number of talks concerned massive binaries: Alberto Sesana talked about ``Probing massive black hole binaries with eLISA and PTA'', and Sean McWilliams on ``Implications of a LISA redesign for massive black hole binary observations''. ``Observational signatures of massive black hole binaries'' were discussed by Carmen Montuori, ``Dual and binary AGN in the cosmic landscape'' by Monica Colpi, ``Massive black hole binaries in gas rich environments'' by Massimo Dotti, and ``Tidal Disruption Flares as a Signature of Supermassive Black Hole Recoil'' by Nick Stone.
 The “Gravitational wave background from binary systems” was discussed by Pablo Rosado; and IMBHs were discussed by Simos Konstantinidis in ``Young clusters and massive BHs as the building blocks of ultra-compact dwarf galaxies''. 

A number of talks concerned our own galactic center: Stefan Gillessen gave an observational status talk, Nadeen Sabha and Mohammad Zamaninasab talked about ``The Galactic Black Hole and its Variable Emission''.  Tal Alexander connected with EMRIs in ``Stellar dynamics near massive black holes: the EMRI / SgrA* connection'', and Cole Miller talked about binary stars in galactic centers in general. 
Tamara Bogdanovi{\'c} suggested that the paradigm of the Milky Way as an inactive galaxy could be forever changed, when she  asked ``Can a Satellite Galaxy Merger Explain the Active Past of the Galactic Center?'' Tamara discussed different pieces of forensic evidence for active past of our galactic center: Fermi bubbles, the X-ray echo of a past luminosity outburst, current star formation, and a deficit of old stars in the galactic center. Some or all of these events may have a common cause: In Tamara's satellite merger scenario,  at $z \approx 10$ a primordial satellite galaxy with an IMBH would have started to merge with the young milky way. A tidally stripped satellite then sinks toward the galactic center, and its central IMBH subsequently merged with the central SMBH less than ten million years ago.

Three talks concerned waveform modeling and solving the Einstein equations: Sascha Husa summarized the current status of the ``Numerical Modeling of Gravitational Waves from Black Hole Binaries'', Niels Warburton and Sarp Akcay presented ``Full self-force EMRI waveforms (for nonrotating MBHs)'' in a twin talk, and Pedro Montero talked about ``Relativistic collapse and explosion of rotating supermassive stars with thermonuclear effects''. In the last regular talk of the program -  ``Taking Advantage of Serendipity'' - Sam Finn talked about what we can learn from GW observations without imposing a particular source model.

The workshop concluded on Friday afternoon with summary talks of the discussion sessions, collecting some of the main points made during the week. 
The shorter baseline reduces sensitivity of eLISA with respect to LISA, and shifts the sensitivity curve of eLISA to higher frequencies/lower masses, with a peak sensitivity a few $10^5$ solar masses. Lighter binaries are thus becoming important, and one of the astrophysical new questions raised by the LISA descoping is whether MBHSs exist with masses below a million solar masses. Another recurring theme was that
 with reduced sensitivity we need to exploit all the waveform features: merger and ringdown in addition to inspiral, spin precession, higher harmonics and eccentricity. There is hope that this increase in sophistication of waveform models with respect to what has been mostly used to formulate LISA science goals and capabilities will compensate for some of the sensitivity loss. Emphasis was put on the fact that full, accurate waveform models based on combining numerical relativity and post-Newtonian/EOB models are yet to be constructed for large portions of the parameter space.

Another question discussed was what would be the most desirable design improvement over the current new baseline - longer arms, longer mission duration, or 6 links? For EMRIs longer arms is more important (Gair). Event rates for EMRIs are still highly uncertain, and dynamical mechanisms for EMRI formation constitute a moving frontier. In any case, a SNR of more than a few hundred. Both Leor Barack and Jonathan Thornburg agreed that general self-force waveforms for  Kerr and including second order corrections should be ready in 10 years.
An important questions concerns the required accuracy of observations and waveform models.  Any intrinsic mass measurements at the 20\% level would be highly desirable, and sky localization $<1$degree combined with distance $<$few\% would be required to have a realistic chance to get a counterpart and (possibly) do cosmology. Distance measurement is not strictly necessary for model selection, since models are in general distinguishable in the redshifted mass-mass ratio-spin parameter space. However, stating that a merging binary has been observed at $z=10$ would be much more convincing for the astrophysics  community, than only being able to claim $z>2$! Spin measurements at the 0.1\% precision would be great to probe maximal MBH spins (0.93? 0.97? 0.998?) and to get insights in the accretion physics. Spin measurements will also be critical to assess the dynamics of MBHBs: Spin alignment will help to discriminate dry or wet coalescence, and spin magnitude will tell us about the typical MBH accretion channels. Measuring spins to 10\% level could already provide important insight on these questions. Eccentricity is expected to be usually $<0.1$ in the eLISA band, but-extremely eccentric binaries might be a smoking gun of some unusual dynamical mechanism.
A largely unresolved questions is how and to which degree we will be able to test GR and fundamental physics, and how useful observations of massive BH binaries will be for this purpose. In principle, if we measure three ringdown modes we can test the no hair theorem, but we need high SNR in at least three modes, and we need to disentangle the single modes. In many cases we lack waveforms for alternative theories. An interesting aspect would be to measure Kerr parameters larger than the extreme Kerr limit of unity. However in this case, one would likely worry about the astrophysical environment and the data analysis before looking for deviations from GR.

It has been heard that many of the participants quite enjoyed the varied cuisine and night life of Palma city, and some stayed on for a few days of holidays on the sunny island -- not just to visit the beaches, but also for Mallorca's
 rock climbing. The next meeting wil be held in September 2012 in Beijing \cite{Beijing_web}.

\end{document}